\documentclass[aps,prb,twocolumn,showpacs,groupedaddress]{revtex4}
\usepackage{graphics,epsfig}
\begin{document}
\title{Electronic polarity of nanoclusters: quantum and many-body 
effects}
\author{A. V. Shytov}
\affiliation{Physics Department, Brookhaven National Laboratory, 
Upton, New York 11973-5000}
\author{P. B. Allen}
\affiliation{Department of Physics and Astronomy,
             State University of New York, Stony Brook, NY 11794-3800
	    }
\altaffiliation{Center for Functional Nanomaterials, Brookhaven National
Laboratory, Upton, NY 11973-5000}
\date{\today}
\begin{abstract}
Interesting electrical polarity in nanoclusters usually requires 
the polarizability to exceed the value $R^3$ of the classical 
sphere of radius $R$. We clarify how this occurs naturally
in single electron quantum systems, and relate it to the giant
polarizability of $Na_{14}F_{13}$, and to spontaneous dipole
formation on niobium clusters. Many-body effects generally
reduce the polarizability through screening. The usual RPA treatment
retrieves the classical answer, but it significantly
overestimates screening in few-electron systems. The system of two 
electrons on the surface of a sphere is solved numerically, 
to account for the Coulomb repulsion. At high densities, 
numerical results agree with RPA model with properly subtracted
self-interaction effects. At low densities, the system 
performs quantum oscillations around the classical ground state. 
We calculate the lowest anharmonic correction
to the polarizability, which also agrees well with numerical evaluation
of the polarizability. 
\end{abstract}
\maketitle

{\it Introduction.} 
Electrical polarity~\cite{Bottcher,Mahan} (both permanent moments $\mu_0$
and induced moments $\mu_{\rm ind}=\alpha F_{\rm ext}$)
are important in determining 
the interaction of a nanocluster
with its environment~\cite{env}  and with external probes such
as light \cite{light}. Large polarizability may also lead to 
spontaneous symmetry-breaking, analogous to bulk
ferroelectricity \cite{Landman,deHeer,Rudiger}.  
Despite good numerical progress 
on the polarizability \cite{Maroulis} 
of small systems, 
its general understanding is not developed. 
In classical electrostatics, 
a metallic sphere of radius
$R$ placed in an external electric field $\vec{F}$
develops surface charge whose field exactly cancels the external field 
in the interior.  The dipole $\vec{\mu}=\alpha\vec{F}$
associated with this surface charge gives the
polarizability of a classical metal sphere, 
$\alpha_{\rm CMS}=R^3$.  Experiments
on various metallic clusters \cite{Kresin} and on
C$_{60}$ \cite{c60} yield values of $\alpha$
in rough agreement \cite{Kresin2}.  However, larger 
$\alpha$ is necessary  for spontaneous dipole formation. 
For example, in a crystal of nearly touching point-polarizable spheres,
the Clausius-Mossotti
polarization catastrophe occurs when $4\pi n\alpha/3 > 1$, where
$n$ is the concentration.  This translates to a
critical polarizability of $\alpha_c=3\sqrt{2}R^3/\pi$, larger
than the available metallic value.

How could larger-than-classical values of $\alpha$ emerge
in metallic clusters?  An analog case is
the prediction \cite{Landman}
and apparent confirmation \cite{xxx} that the
system Na$_{n+1}$Cl$_n$ (and especially the highly
symmetric Na$_{14}$Cl$_{13}$ case), with one electron 
in a loosely-bound ``surface state'' outside a
closed shell, has a second-order Jahn-Teller instability
to a polar state.  This suggests that the total polarizability
($\alpha_{\rm el}=$ electronic plus $\sum Z_{\rm eff}^2/\omega^2=$
vibrational, with $Z_{\rm eff}$ being appropriate
Born effective charges and $\omega$ the relevant
vibrational frequencies \cite{Baruah}) has diverged.
Similarly, de Heer's group \cite{deHeer}
found low $T$ permanent dipole moments
on Nb clusters which might be interpreted
as $\alpha(T)$ increasing to a divergence as $T$ decreases.

In this Letter, we analyze the polarizability of a nanocluster
with a surface state occupied by an electron. As a  model for this
system, we consider an electron confined to the surface of a sphere. 
We show that due to quantum effects, 
electronic polarizability can considerably exceed 
the metallic value $\alpha_{CMS}$ at low temperatures. 
This result implies that the coupling between the surface  electron 
and elastic modes of nuclear displacement can
decrease the vibrational frequencies and drive the
total polarizability to diverge.  
We also analyze how the classical value $\alpha_{\rm CMS}$ is recovered by
screening when the number of metallic electrons increases.
In particular, we consider the case of two surface electrons
which is realized in the neutral cluster Mg$_{14}$O$_{13}$.
We predict that when a second electron is added,
the polarizability diminishes to
values closer to $\alpha_{\rm CMS}$, and
stability of the symmetric structure is restored.

{\it Polarizability of one electron on the sphere.}
Confining electrons to the surface of a sphere permits
theoretical simplification \cite{Tempere}.
This model is not as much of a ``spherical cow'' \cite{cow}
as might be thought.  For example, the C$_{60}$ molecule confines 60
carbon $\pi$ electrons to the region near the surface of a
sphere.  In neutral undistorted Na$_{14}$Cl$_{13}$, 
one outer electron lies in a shell similar to a sphere.
``Core-shell'' nanoparticles with insulating cores and metallic
shells have been studied \cite{Averitt}.
The electron eigenstates are $\psi_L =Y_L (\Omega)$, the spherical
harmonics, with $L=(\ell,m)$, and energies $E_L=(\hbar^2/2mR^2)\ell(\ell+1)$.
The polarizability $\alpha=\langle\mu_z \rangle /F_z$ is
\begin{equation}
\alpha(T)=2\sum_{L^{}L^{\prime}} w_L 
\frac{|\langle L^{\prime}|\mu_z |L\rangle|^2}{E_L-E_L^{\prime}}
\label{eq:pol}
\end{equation}
where the
dipole operator $\mu_z$ is $ -eR\cos\theta$ and
$w_L=w_{\ell}=\exp(- E_L /k_B T)/Z$ is the probability that the
system is in state $L$.  At $T=0$, the electron is in the
ground state ($w_L=\delta_{L,0}$), and
there is only one non-zero matrix element,
$\langle 1,0|\cos\theta|0,0\rangle=1/\sqrt{3}$.
Then we get
\begin{equation}
\alpha(T=0)=\frac{2R^4}{3a_B}
\label{eq:polqu}
\end{equation}
where $a_B$ is the Bohr radius.  This result is interesting, because
it shows that for unscreened one-electron
systems, the value of $\alpha$ can exceed
the value $\alpha_{CMS}=R^3$.
The scaling with $R^4/a_B$ is not restricted to the
spherical shell, but is general for finite system with one electron,
{\it e.g.}, a cubic box.  The high temperature ($k_B T \gg \hbar^2/mR^2$)
answer can be found from classical statistical mechanics, 
and is given by Debye-Langevin law\cite{Debye}:
\begin{equation}
\alpha({\rm high} \ T)=\frac{e^2 R^2}{3k_B T}.
\label{eq:polcl}
\end{equation}
The crossover from the $T=0$ quantum answer to the classical
answer occurs at crossover temperature
$k_B T_\ast \approx \hbar^2/2mR^2$.  
\begin{figure}
\label{fig-alpha-invT}
\centerline{\scalebox{0.32}[0.32]{\includegraphics{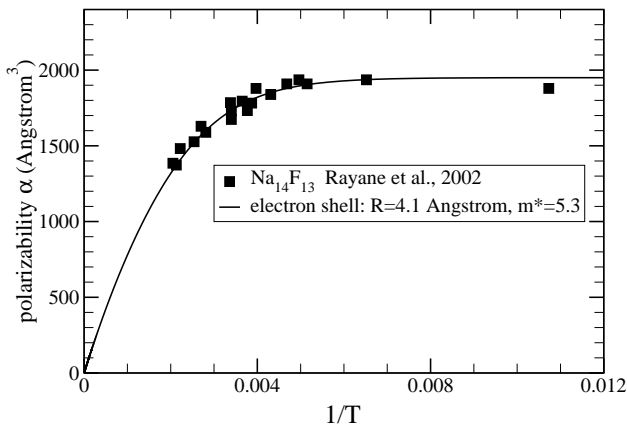}}}
\caption{\label{fig:x} Polarizability versus inverse temperature in K
for Na$_{14}$F$_{13}$ clusters from Ref. \onlinecite{Rayane}, 
compared with theory for a single electron on a sphere.}
\end{figure}
Consider the outer electron of
Na$_{14}$F$_{13}$,
responsible for the giant
polarizability measured by Rayane {\it et al.} \cite{Rayane}.
The $T$-dependence can be interpreted as
a large low $T$ value from Eq. (\ref{eq:polqu}), which
evolves at higher $T$ toward the classical value of Eq. (\ref{eq:polcl}).
At intermediate temperatures, the polarizability 
can be easily tabulated numerically.  
Using the well-known matrix elements of a dipole moment \cite{Condon}
\begin{equation}
\left\langle l + 1, m | \cos\theta | l, m \right\rangle
=i \sqrt{\frac{(\ell+1-m)(\ell+1+m)}{(2\ell+1)(2\ell+3)}},
\label{eq:c}
\end{equation}
one can rewrite Eq. (\ref{eq:pol}) as
\begin{eqnarray}
\alpha(T)&=&6\alpha(0)\sum_{\ell} \left(\frac{w_{\ell}-w_{\ell+1}}
{(\ell+1)(\ell+2)-\ell(\ell+1)}\right) \nonumber \\
&\times&\sum_{m=-\ell}^{\ell}
\frac{(\ell+1)^2-m^2}{(2\ell+1)(2\ell+3)}.
\label{eq:sum} 
\end{eqnarray}
The sum over $m$ can now be done, giving
\begin{equation}
\alpha(T)=\alpha(0)\sum_{\ell}(w_{\ell}-w_{\ell+1})=\alpha(0)w_0=\alpha(0)/Z.
\label{eq:alpha}
\end{equation}
The computation reduces to calculation of the
partition function $Z=\sum_{\ell}(2\ell+1)e^{-\ell(\ell+1)/\tau}$, 
with the dimensionless temperature $\tau = T / T_\ast$. 
Figure 1 shows values of $\alpha(T)$ 
from Eq.~(\ref{eq:alpha}) plotted versus $1/\tau$ and
compared with the data \cite{Rayane} for Na$_{14}$F$_{13}$.
Two fitting parameters were used: the value $\alpha(0)$ was 
taken to be 1950 \AA$^3$, and the crossover temperature 
$T_\ast = \hbar^2/2mR^2 = 975 {\rm K}$. 
Using Eq. (\ref{eq:polqu}), 
one finds reasonable values, $R=$ 4.1 \AA  \ and
an effective mass $m^{\ast}=$ 5.3 free electron masses.  
(Of course, rotational and vibrational sources of $T$-dependence
of $\alpha$ are probably not negligible, but the limited experimental
range of $T$ does not warrant a closer fit.) For comparison, 
based on the Na-F bond length $a = 2.3$\AA, one can estimate the
cluster size to be about $R$=4\AA. Note that the low temperature
polarizability is much larger than $\alpha_{CMS} = 70$\AA$^3$. 

One can consider an alternative explanation of the $1/T$ behaviour
of the susceptibility. One can imagine that the cluster in fact
has a large spontaneous dipole moment with fixed orientation
with respect to cluster. Thermal rotations in the absense
of an external electric field would randomize the direction
of polarization, and an external field would align the dipole. 
Any mechanism of this sort, however, would produce saturation
temperature much larger than $T_{\ast}$ because of a large mass
associated with dipole rotations. This would be clearly
incompatible with saturation temperature observed in Ref.~\onlinecite{Rayane}.

In our calculation, we assumed that the response is linear in 
the applied field. For this assumption to be valid, 
the  polarization energy $\alpha {\cal E}^2/2$
must be smaller than the level spacing $\hbar^2/mR^2$. 
For larger fields, the dipole moment would saturate. 
One can estimate the saturation field as 
${\cal E}_{\rm sat} = {\cal E}_{\rm at} (a_B/R)^3$, 
where ${\cal E}_{\rm at} = e / a_B^2$ is atomic electric 
field, which is of the order of $10^{10}$ V/m. One can see from
this estimate that although the saturation field quickly decays
with the cluster radius, it is still rather large for reasonable
values of $R$.

{\it Screening effects.}
If there are many electrons on the shell, 
a mean field (RPA or time-dependent Hartree)
approximation may be reasonable.  
In \cite{Tempere, Bertsch}, polarizability was derived via RPA. 
Below we give a simple derivation and discuss the effects due
to finite number of electrons. 
Consider the metallic sphere containing $N$ electrons 
in the external field ${\cal E}$. The electric potential 
on the surface of the sphere is generated both by the external 
electric field source and electrons. In polar coordinates, 
the external field contribution is $-{\cal E} r \cos\theta$, 
while the dipole moment contribution is $\mu_z \cos\theta / r^2$
for $r\geq R$. 
In the mean field theory, electrons respond to the 
screened potential on the surface $r=R$, 
which is proportional to the effective field 
${\cal E}_{\rm eff} = {\cal E} - \mu_z / R^3$. The self-consistency 
condition is then
\begin{equation}
\label{eq:selfcons}
\mu_z = \alpha_0 {\cal E}_{\rm eff} 
      = \alpha_0 \left({\cal E} - \frac{\mu_z}{R^3}\right)
\ . 
\end{equation}
Solving Eq.~(\ref{eq:selfcons}) for $\mu_z$, one arrives at
\begin{equation}
\alpha_{\rm MF}=\frac{\alpha_0}{1+\alpha_0/R^3}.
\label{eq:alphasc}
\end{equation}
This calculation overestimates screening effects: the actual
field acting on an electron should not include the field created
by the electron itself. Assuming that all electrons contribute equally
to the screening field, 
one can exclude self-interaction by multiplying the screening field  
$-\mu_z / R^3$ by $(1 - 1/N)$. The result
\begin{equation}
\alpha_{\rm MFC}=\frac{\alpha_0}{1+(1-1/N)\alpha_0/R^3},
\label{eq:alphasccorr}
\end{equation}
unlike Eq.~(\ref{eq:alphasc}), behaves correctly at $N = 1$. 

A realization of a dense metallic spherical shell
is the $\pi$-electron system of the C$_{60}$ molecule
\cite{Bertsch}.  Pederson
and Quong \cite{Pederson} computed the polarizability of C$_{60}$
both without and with self-consistent screening.  Their answer
for $\alpha_0$ is 35 $a_B^3$ per C atom, while after screening
they find $\alpha_{\rm MF}$ to be 9.3 $a_B^3$ per C atom.
The electrons are on a spherical shell which is more
than infinitesimally thin, but let us use the thin model
anyway.  To account for the reduction by screening, we need
$R=9.0a_B$.  This compares reasonably with the radius 6.7-7.0 $a_B$
of the carbon nuclei, adding an extra Angstrom of distance for
the finite extent of the $\pi$ electron system.

\begin{figure}
\label{fig-alpha-2-log}
\centerline{\scalebox{0.32}[0.32]{\includegraphics{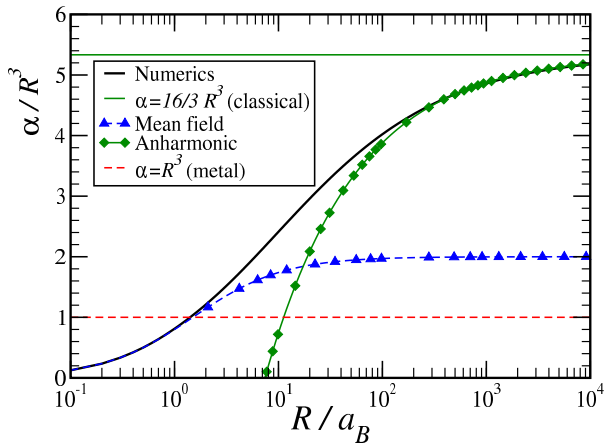}}}
\caption{Polarizability of two electrons on a sphere vs. its radius $R$, 
normalized to 
the classical value $R^3$ on a logarithmic scale. The mean field
(Eq.~(\ref{eq:alphasccorr}) and semiclassical (Eq.~(\ref{alpha-anharmonic})) 
results are also shown for comparison.} 
\label{fig-alpha-cap}
\end{figure}

{\it Two-electron model.}
The Coulomb interaction between electrons leads to correlation
between their positions, which is omitted in mean field
theory. 
To clarify the role of correlation, we
consider two electrons on the sphere.  
Related studies of few-electron model systems
exist \cite{Yannouleas,Thompson2} 
but polarizability was not analyzed.  We use the Hamiltonian
\begin{equation}
\label{hamiltonian-interacting}
H = - \frac{\hbar^2}{2 m R^2} \left(\nabla_1^2 + \nabla_2^2\right) 
+ \frac{e^2}{R |{\bf n}_1 - {\bf n}_2|}
\ ,
\end{equation}
where ${\bf n}_{1, 2}$ are positions of the electrons on the sphere 
(${\bf n}_{1,2}^2 = 1$), and $\nabla_{1,2}^2$ are 
spherical Laplace operators. The first term 
in Eq.~(\ref{hamiltonian-interacting}) represents the kinetic energy, 
and the second term describes Coulomb repulsion between electrons. 
The polarizability at $T = 0$ is again given by Eq.~(\ref{eq:pol}), 
with  $|L\rangle$ being exact two-particle states, $E_L$ their energies, 
and  $\mu_z = e R (n_{z1} + n_{z2})$ being total dipole moment. 
There are multiple states with a given total angular
momentum $L'$.  The  sum in Eq.~(\ref{eq:pol}) is taken over all 
such states. 

Solving for eigenstates of the Hamiltonian 
(\ref{hamiltonian-interacting}) is difficult even at zero temperature
because of a vast configuration space. 
To reduce its dimension, we use symmetry considerations. 
The  ground state $\Psi_0({\bf n}_1, {\bf n}_2)$ 
is symmetric with respect 
to electron coordinates and has total angular momentum $\ell = 0$.
Thus, the mass center of the two-electron system is completely 
delocalized, 
and $\Psi_0$ only depends on the distance between the electrons. 
We use the spherical angle $\gamma$  to parametrize this distance 
($\cos \gamma = {\bf n}_1 \cdot {\bf n}_2$), so the wave function 
takes the form
$\Psi_0 ({\bf n}_1, {\bf n}_2) = \Psi_0 (\gamma)$.
To compute the action of the kinetic energy term $\nabla_{1,2}^2$ in  
Eq.~(\ref{hamiltonian-interacting}) on $\Psi_0 ({\bf n}_1, {\bf n}_2)$, 
it is convenient to use the following trick. 
Note that the kinetic energy in the $\ell = 0$ state 
also depends on the relative angle $\gamma$ only, 
and therefore one can compute it at some particular values of ${\bf n}_{1,2}$
and extend the result for all ${\bf n}_{1,2}$ with the same value of $\gamma$.
A convenient point corresponds to the second electron at the north pole.
Then, the angle $\gamma$ 
equals the polar angle of the first electron, and the kinetic energy
is given by the well-known expression for the Laplace operator on the sphere. 
Treating the second electron in the same manner, one obtains the
Schr\"{o}dinger equation for $\ell = 0$ states:
\begin{equation}
\label{ground-state-0}
\left\{- \frac{\hbar^2}{m R^2} \frac{1}{\sin \gamma} 
             \frac{\partial}{\partial \gamma} 
             \left(\sin \gamma \frac{\partial}{\partial \gamma} \right) 
            + U_{\ell = 0} (\gamma)
           \right\} \Psi_0 (\gamma) = E_0 \Psi_0 (\gamma)\ , 
\end{equation}
where the effective potential 
\begin{equation} 
\label{u-0}
        U_{\ell = 0} (\gamma) = \frac{e^2}{2R \sin\frac{\gamma}{2}}
\end{equation} 
is entirely due to the Coulomb interaction.
 
Now consider excited states. The only non-zero matrix elements in 
Eq.~(\ref{eq:pol}) 
are between $\ell = 0$ and $\ell = 1$
states. Moreover, since  both $\Psi_0$ and $\mu_z$ are symmetric with respect
to permutation of the electrons, one has to consider only those
$\ell = 1$ states in which electron momenta are added symmetrically. 
Up to a scalar multiplier, one can construct only one symmetric vector
out of ${\bf n}_{1,2}$: 
${\bf N} = ({\bf n_1} + {\bf n}_2) / (2 \cos\frac{\gamma}{2})$
(we normalized ${\bf N}$ so that ${\bf N}^2 = 1$).
In other words, the transformation properties of $\ell = 1$ states
are identical to those of the mass center of the system. 
The wave function of the states with $\ell = 1$, $\ell_z = 0$  therefore
can be written in the form
$\Psi_1^{(n)} ({\bf n}_1, {\bf n}_2) = N_z \Psi_1^{(n)} (\gamma)$.
By a direct calculation, we find  that the Schr\"{o}dinger 
equation for $\Psi_{1}(\gamma)$ has a form similar to 
Eq.~(\ref{ground-state-0}) with a different effective potential:
\begin{equation}
\label{u-1}
U_{\ell = 1} (\gamma) = U_{\ell = 0} (\gamma) + \frac{\hbar^2}{4mR^2} 
               \left(
                        1 + \frac{1}{\cos^2\frac{\gamma}{2}}
               \right)
\ . 
\end{equation}
The last contribution in Eq.~(\ref{u-1}) is the kinetic energy 
of joint rotation of the two electrons with angular momentum $L = 1$.
One can derive this ``centripetal'' term by a different method. 
At fixed distance between the electrons $\gamma$, 
the rotational dynamics of the two-electron system 
can be described as that of a quantum asymmetric top with moments of inertia
$I_1 = 2 mR^2 \sin^2\gamma/2$, $I_2 = 2 m R^2$,
and $I_3 = 2 m R^2 \sin^2\gamma/2$. 
It is known (see, {\it e.g.}, \cite{LL-top}) that $\ell = 1$ is a special 
case in the quantum top problem: the wavefunctions of these states
are completely determined by the symmetry and therefore independent
of $I_{1,2,3}$. 
In our case it means that the top dynamics is independent  
of  oscillations of $\gamma$. 
This allows one to separate the variables for $\ell = 1$ state. 
The wave function
of the two electron system
can be written as $\Psi_{1}({\bf n}_1, {\bf n}_2) 
= \Psi_T (\xi, \eta, \zeta) \Psi_\gamma (\gamma)$, where 
$\Psi_T$ is the wavefunction of the top described by Euler angles
$\xi$, $\eta$ and $\zeta$.  (One can also check that this top 
wavefunction is proportional to ${\bf N}$.)
For symmetric $\ell = 1$ states with zero
projection of $J$ on the top axis, the energy of the top is
$\frac{\hbar^2}{2} (1/I_1 + 1/I_2)$, which coincides with the last 
term in Eq.~(\ref{u-1}).

One can now use Eqs.~(\ref{ground-state-0})-(\ref{u-1})  
to find eigenfunctions and eigenvalues by solving a one-dimensional
boundary-value problem. (The wave function has to be regular everywhere
on the sphere. However, a general solution to Eq.~(\ref{ground-state-0})
may have singularities at $\gamma = 0$ and $\gamma = \pi$, where
the coefficients of Schr\"{o}dinger equation are singular.
Thus, the requirement of regularity at these points serves as two
independent boundary conditions.) 
While in the limit of 
$R \ll a_B$ one can thus reproduce the mean field theory 
result~(\ref{eq:alphasccorr}), the opposite limit of strong 
interaction ($R \gg a_B$) is more interesting. In this regime, 
Coulomb repulsion dominates over kinetic energy. 
In the ground state, electrons
perform zero-point oscillations at opposite poles of the 
sphere, thus minimizing potential energy. 
The electric field shifts the minimum and changes the energy
of the system. The extra energy due to the field is  
$U_{\cal E}(\gamma) =  2 e{\cal E}_\perp R\cos \gamma/2$, where
${\cal E}_\perp$ is the field component perpendicular to the
axis connecting electrons. Minimizing 
the potential energy $U_{\ell = 0}(\gamma) + U_{\cal E}(\gamma)$, one finds
the ground state energy to be $e^2/2R - 4 R^3 {\cal E}_{\perp}^{2}$. 
Averaging over all directions of the axis, one can replace 
${\cal E}_{\perp}^2$ by $2{\cal E}^2/3$. Thus, the polarizability of the 
two-electron system in the classical limit is $16 R^3/3$, 
still higher than the metallic value. 

However, it turns out that the amplitude of zero point motion around the 
equilibrium positions decays only as $(a_B/R)^{1/4}$, and 
the simple classical picture described above holds only 
for impractically large values $R /a_B > 10^4$. To improve the
classical approximation, one can expand Eqs.~(\ref{ground-state-0})-(\ref{u-1})
in the vicinity of the classical equilibrium, $\gamma \approx \pi$.
For the $\ell = 0$ state, Schr\"{o}dinger equation takes the form: 
\begin{equation}
\frac{\hbar^2}{mR^2}(\hat{h}_{0} + \delta \hat{h}_{0}) \Psi_0 (x) 
= E_0 \Psi_0 (x)
\ , 
\end{equation}
where $x = \pi - \gamma$ is the deviation from equilibrium. 
The dimensionless Hamiltonian 
\begin{equation}
\hat{h}_0 = - \frac{\partial^2}{\partial x^2} 
            - \frac{1}{x} \frac{\partial}{\partial x}
	    + \frac{R}{2a_B} \left( 1 + \frac{x^2}{2}\right)
\end{equation}
describes harmonic oscillations near the equilibrium,
and 
\begin{equation}
\delta\hat{h}_0 = \frac{x}{3} \frac{\partial}{\partial x} 
                + \frac{5}{3\cdot128} \frac{R}{a_B} x^4
\end{equation}
is the first anharmonic correction to $\hat{h}_0$. 
For $\ell = 1$, one finds 
a similar equation with harmonic and anharmonic parts given by 
\begin{equation}
\hat{h}_1 = \hat{h}_0 + \frac{1}{x^2} \ ; \qquad 
\delta\hat{h}_1 = \delta\hat{h}_0 + \frac{1}{3}\ . 
\end{equation}
If one ignores anharmonic parts, these equations can be satisfied
by wave functions of a harmonic oscillator: 
\begin{eqnarray}
\psi_0(x) &=& \exp (-x^2 / 2 x_0^2) 
\\
\psi_1(x) &=& x \psi_0 (x)
\end{eqnarray}
where 
\begin{equation}
      x_0 = (4a_B/R)^{1/4}
\end{equation}
is the amplitude of zero point oscillations.
The corresponding eigenvalues are 
\begin{equation}
\label{eq:eharm}
E_0 = \frac{e^2}{2R} \left(1 + \sqrt{\frac{a_B}{R}}\right) \ ; \qquad 
E_1 = \frac{e^2}{2R} \left(1 + 2 \sqrt{\frac{a_B}{R}}\right)
\ .
\end{equation}
The first term $e^2/2R$ is the classical Coulomb energy, and the second term
is due to quantum oscillations. 

The zero point motion amplitude $x_0$ can be considered as
a small anharmonicity parameter. We shall now compute 
anharmonic corrections to the polarizability, expanding 
over the powers of $x_0^2$. 
Treating anharmonic terms $\delta h_0$ and $\delta h_1$
by standard perturbation theory, 
one can compute the corrections to the energy levels:
\begin{eqnarray}
\label{eq:eanharm}
\delta E_0 &=& \langle \psi_0 | \delta \hat{h}_0 | \psi_0 \rangle = 
- \frac{e^2}{R}\frac{x_0^4}{16}
\\ 
\delta E_1 &=& \langle \psi_1 | \delta \hat{h}_1 | \psi_1 \rangle =
\frac{e^2}{R}\frac{5x_0^4}{16}
\ .
\nonumber
\end{eqnarray}
Similarly, one can also find corrections to the wave functions 
\begin{eqnarray}
\label{eq:psicorr}
\delta \psi_0 &=& \frac{x_0^2}{2} 
                  \left(
		      - \frac{\xi^4}{24} - \frac{\xi^2}{9} - \frac{1}{36}
		  \right) e^{-\xi^2/2}
\\
\delta \psi_1 &=& \frac{x_0^2}{2} 
                  \left(
		      \frac{5 \xi^5}{144} - \frac{\xi^3}{16} - \frac{\xi}{12}
		  \right) e^{-\xi^2/2}
\ ,
\nonumber
\end{eqnarray}
where $\xi = x/x_0$.

One can now use $\delta E_{0, 1}$ and $\delta \psi_{0, 1}$ to 
calculate the correction to the polarizability given by Eq.~(\ref{eq:pol}).
To do that, one has to take into account
anharmonic corrections to the dipole moment operator:
\begin{equation}
\label{eq:dipolemoment}
{\bf \mu} = 2 eR {\bf N}
    \cos \frac{\gamma}{2} \approx 2e R {\bf N} 
    x_0 \xi \left( 1 - \frac{\xi^2 x_0^2}{24}\right)
\end{equation}
(here ${\bf N}$ is the  unit vector directed to the mass center),
and to the volume element in the integral:
\begin{equation}
d\Omega \propto \sin \gamma d\gamma \approx 
          x_0^2 \left( 1 - \frac{\xi^2 x_0^2}{6}\right) \xi d\xi
\ .
\end{equation}
(One can check that this volume element is consistent with kinetic part
of the Hamiltonian.) In the lowest order in anharmonic corrections, 
the only important contribution comes from the lowest $\ell = 1$ state.
The contributions of higher $\ell = 1$ states are absent in harmonic
approximation, and their contributions to the matrix element
are proportional to anharmonicity parameter $x_0^2$. 
Since the polarization contains square of the matrix element, 
excited $J=1$ states contribute only in the fourth order in $x_0$, 
which we do not consider here. 
Substituting Eq.~(\ref{eq:dipolemoment}) into Eq.~(\ref{eq:pol})
and replacing the matrix
element of $N_z$ between $J=0$ and $J=1$ states by $1/\sqrt{3}$, 
one finds:
\begin{equation}
\label{eq:polar-dimensionless}
\alpha 
= \frac{2 e^2 R^2}{3} \frac{x_0^2}{E_1 + \delta E_1 - E_0 - \delta E_0}
\frac{M^2}{N_0 N_1}
\ , 
\end{equation}
where 
\begin{equation}
M = 2\int\limits_{0}^{\infty} \xi^2 d\xi \left(1 - \frac{5 \xi^2 x_0^2}{24}\right)
(\psi_0 + \delta \psi_0) (\psi_1 + \delta \psi_1)
\end{equation}
and $N_{0, 1}$ are normalization integrals, with corrected volume element:
\begin{equation}
N_i = 2\int\limits_{0}^{\infty} \left( 1 - \frac{\xi^2 x_0^2}{6}\right) 
(\psi_i + \delta\psi_i)^2 \xi d\xi
\ .
\end{equation}
Substituting wave function corrections~(\ref{eq:psicorr}), 
one finds, in the lowest order:
\begin{eqnarray}
\label{eq:elements}
N_0 &=& 1 - \frac{7x_0^2}{18} \ ; \qquad 
N_1 = 1 - \frac{x_0^2}{3} \ ; \\
M   &=& 1 - \frac{2x_0^2}{3} 
\ .
\nonumber
\end{eqnarray}
The energy difference, according to Eqs.~(\ref{eq:eharm}) 
and~(\ref{eq:eanharm}), is
\begin{equation}
\label{eq:energy}
E_1 - E_0 + \delta E_1 - \delta E_0 = \frac{e^2}{R} 
\left(\frac{x_0^2}{4} + \frac{3 x_0^4}{8} \right)
\end{equation}
Substituting Eqs.~(\ref{eq:elements}) and~(\ref{eq:energy})
into Eq.~(\ref{eq:polar-dimensionless}), 
one finally obtains:
\begin{eqnarray}
\label{alpha-anharmonic}
\alpha_{2} (R \gg a_B) &\approx& \frac{ 16 R^3}{3} 
\left( 1 - \frac{49}{36}x_0^2\right) 
\\
&=& 
\frac{ 16 R^3}{3}
\left[ 1 - \frac{49}{18} \left(\frac{a_B}{R}\right)^{1/2} 
   \right] 
\ . 
\nonumber
\end{eqnarray}

The results of numerical solution of the two-electron problem 
are shown on Fig.~\ref{fig-alpha-cap}. 
The self-consistent result~(\ref{eq:alphasccorr}) 
breaks down at $R \sim 3 a_B$, while the classical 
regime~(\ref{alpha-anharmonic}) takes over at $R \geq 100 a_B$. 
For reasonable size nanoclusters ($R \sim 10 a_B$), 
the effect of electron correlations is significant, but the polarizability
is still considerably higher than the classical value $\alpha = R^3$. 

The limit $R < a_B$ shown in the left part of the plot may seem unphysical. 
However, note 
that the Bohr radius entering our calculation is in fact an 
effective parameter, depending on the effective mass of the surface 
state. If, for some reasons, the effective state is a light electron, 
the effective Bohr radius can also be large, and the limit $R < a_B$
may become physically applicable. 

It is also instructive to check whether the Clausius-Mossotti catastrophe
($4\pi/3 n \alpha >  1$) can be achieved. 
Packing spheres in the most dense way, 
one requires $\alpha/R^3 > 3\sqrt{2}/\pi $, which is achieved 
for $R \geq  2.5 a_B $.
Thus, despite screening, an array of two-electron nanoclusters 
can develop a spontaneous polarization. 

To summarize, the polarizability of a nanocluster with a surface electron 
state may be significantly  enhanced due to quantum effects. This
enhancement, however, disappears when the number of ``metallic'' electrons
is increased. Although we considered a very simple model of a nanocluster, 
our results should hold qualitatively for more complex clusters, 
including niobium ferroelectric clusters studied in Ref.~\onlinecite{deHeer}.  
At the first sight, these clusters do not look like a single-electron systems.
However, if the spectrum of the high $T$ state has its
outermost electron in a separated level with orbital
degeneracy, as in Na$_3$ (see Ref.\onlinecite{Allen}),
or with quasi-Jahn-Teller degeneracy as in Na$_{14}$Cl$_{13}$,
similar physics probably applies.
Also, despite the fact that we only considered surface electrons, 
one can expect a similar physics when electrons are confined in 
the bulk of the cluster. For example, the estimate for the level
spacing, dipole moment and therefore for polarizability,
should be still valid for bulk electrons. Thus, our results
can give correct qualitative description for more complex
clusters with a small number of ``mobile'' electrons.

We thank A. G. Abanov, 
A. Durst, J. T. Muckerman, and M. R. Pederson for helpful
discussions. Research at Brookhaven National Laboratory
was supported by U.S. DOE under
contract No. DEAC 02-98 CH 10886. 
Work at Stony Brook  was supported in part by NSF grant no.
NIRT-0304122. 

\end{document}